\documentclass[11pt]{article}
\usepackage[margin=2.5cm, top=2.5cm, headsep=1.5cm]{geometry}
\usepackage{graphicx}%
\usepackage{multirow}%
\usepackage{amsmath,amssymb,amsfonts}%
\usepackage{amsthm}%
\usepackage{mathrsfs}%
\usepackage[title]{appendix}%
\usepackage{xcolor}%
\usepackage{textcomp}%
\usepackage{manyfoot}%
\usepackage{booktabs}%
\usepackage{algorithm}%
\usepackage{algorithmicx}%
\usepackage{algpseudocode}%
\usepackage{listings}%
\usepackage{geometry}
\usepackage{authblk} 
\usepackage{lineno}
\usepackage{titlesec}
\usepackage{hyperref} 
\hypersetup{
    colorlinks=true,
    linkcolor=blue, 
    citecolor=blue, 
    urlcolor=blue   
}
\usepackage[superscript,biblabel]{cite}

\providecommand{\Jms}[0]{$J/m^2$}
\providecommand{\um}[0]{$\mu m$}

\providecommand{\uSENB}[0]{$\mu$SENB}
\providecommand{\rp}[0]{$r_p$}
\providecommand{\lch}[0]{$l_{ch}$}

\providecommand{\SI}[0]{Supplementary Materials}
\providecommand{\DC}[0]{$DC$}
\providecommand{\DClow}[0]{$DC_{low}$}
\providecommand{\DCmed}[0]{$DC_{med}$}
\providecommand{\DChi}[0]{$DC_{high}$}
\providecommand{\DCi}[0]{$DC_{17}$}
\providecommand{\DCj}[0]{$DC_{27}$}
\providecommand{\DCk}[0]{$DC_{80}$}
\providecommand{\Gt}[0]{$G_{t}$}
\providecommand{\Gp}[0]{$G_{p}$}
\providecommand{\Gf}[0]{$G_{f}$}
\providecommand{\Jq}[0]{$J_{Q}$}

\providecommand{\keywords}[1]{\textbf{\textit{Keywords:}} #1}
\begin{document}
    \title{Rethinking Ductility - A Study Into the Size-Affected Fracture of Polymers}
    
    \author[a]{Zainab S. Patel\thanks{Corresponding author: patelz@uw.edu}}
    \author[b]{Abdulaziz O. Alrashed}
    \author[b]{Kush Dwivedi}
    \author[c]{Marco Salviato}
    \author[b]{Lucas R. Meza} 
    
    \affil[a]{Materials Science and Engineering, University of Washington, Seattle, WA, 98195}
    \affil[b]{Mechanical Engineering, University of Washington, Seattle, WA, 98195}
    \affil[c]{Aeronautics and Astronautics, University of Washington, Seattle, WA, 98195}
    \date{}
    \maketitle

    
    \section*{Abstract}
        Ductility quantifies a material's capacity for plastic deformation, and it is a key property for preventing fracture driven failure in engineering parts.
        While some brittle materials exhibit improved ductility at small scales, the processes underlying this phenomenon are not well understood.
        This work establishes a mechanism for the origin of ductility via an investigation of size-affected fracture processes and polymer degree of conversion (DC) in two-photon lithography (TPL) fabricated materials.
        Microscale single edge notch bend (\uSENB) specimens were written with widths from 8 to 26 \um{} and with different laser powers and post-write thermal annealing to control the DC between 17\% and 80\%.
        We find that shifting from low to high DC predictably causes a $\sim$3x and $\sim$4x increase in strength and bending stiffness, respectively, but that there is a corresponding $\sim$6x decrease in fracture energy from 180 $J/m^2$ to 30 $J/m^2$.
        Notably, this reduced fracture energy is accompanied by a ductile-to-brittle transition (DBT) in the failure behavior.
        Using finite element analysis, we demonstrate that the DBT occurs when the fracture yielding zone size (\rp{}) approaches the sample width, corresponding with a known fracture size-affected transition from flaw-based to strength-based failure.
        This finding provides a crucial insight that ductility is a size-induced property that occurs when features are reduced below a characteristic fracture length scale and that strength, stiffness, and toughness alone are insufficient predictors of ductility.
    
    \vspace{5 cm}

    \pagebreak
    \section{Introduction}
        Ductility is an emergent property that quantifies a material's ability to sustain plastic deformation prior to fracture.
        Materials with active plasticity mechanisms, e.g. dislocations in metals and chain sliding in polymers, generally have a higher toughness and are correspondingly thought to be more ductile.
        Brittle materials like ceramics often have higher strength and better corrosion resistance, but their susceptibility to fracture hinders their use as structural materials.
        Strategies for enhancing toughness using extrinsic mechanisms like crack deflection and fiber pullout have long been explored, and with great promise \cite{evans1990perspective, ritchie2011conflicts, mirkhalaf2014overcoming, bouville2014strong}. 
        Combining intrinsic ductility with extrinsic toughening mechanisms holds the potential to yield tough, damage-tolerant components akin to natural materials such as nacre and bone \cite{launey2010mechanistic, launey2010mechanistic-bone, dastjerdi2013weak}.
        Ductile-to-brittle transitions have been extensively observed across metals \cite{li2005ductile, mullner1997ductile}, polymers \cite{brown1982model, quagliato2022quasi}, and ceramics \cite{karch1987ceramics, lawn1994making}, and are generally analyzed according to how composition and microstructure promote plastic deformation.
        Interestingly, some studies have shown that homogeneous materials considered to be intrinsically brittle like silicon \cite{hirsch1989brittle, ostlund2009brittle, issa2021situ}, amorphous carbon \cite{albiez2019size}, and metallic glass \cite{chen2015ductility} can exhibit ductility when made sufficiently small, but there is little mechanistic understanding of this process. 
        
        Discerning the origins of ductility first requires an understanding of fracture processes.
        Prior to fracture, materials will develop a yielding zone in front of a crack which comprises a fracture process zone (FPZ) and a plastic zone (PZ).
        If the sample size is larger than this yielding zone, failure will be fracture-governed, while samples whose size is smaller than this yielding zone will undergo strength-driven failure.
        In the strength-driven regime, materials with a large PZ and small FPZ will experience ductile fracture, while materials with a small FPZ and large PZ will experience quasi-brittle fracture.
        Many studies in this field, pioneered by Ba{\v{z}}ant, have explored the size-affected transition from fracture-driven to strength-driven failure \cite{bavzant1984size, bavzant1990size, bavzant1999size, bavzant2021quasibrittle}, but they often focus on the peak load behavior in quasi-brittle materials like concrete and rock and ignore emergent small-scale ductility.
        
        Two-photon lithography direct laser writing (TPL-DLW) provides an ideal platform for studying size-affected fracture due to its exceptional capability to create parts with feature sizes as small as 100 nm \cite{maruo1997three, lee2008advances}.
        It has been highly successful in creating metamaterials with novel mechanical performance in part because it enables the utilization of size-enhanced nanomaterial properties \cite{meza2014strong, bauer2016approaching, bauer2017nanolattices, xia2022responsive}.
        Significant work has gone toward understanding the role of process parameters on TPL polymer performance, particularly the relationship between the degree of conversion (\DC{}) -- i.e., the extent of cross-linking between polymer chains -- and strength and stiffness \cite{bauer2019programmable, bauer2020thermal}.
        While these properties are important, there is currently a lack of information on how TPL process parameters and the resulting characteristic length scales affect fracture performance.

        In this study, we use microscale single-edge notch bend ($\mu$SENB) experiments to investigate the strength, stiffness, toughness, and ductility of a TPL-DLW polymer as a function of \DC{} and part size.
        Experiments reveal that increasing the \DC{} from 17\% to 80\% increases both strength and bending stiffness by a factor of $\sim$3x and $\sim$4x, respectively, but that there is a corresponding $\sim$6x reduction in the toughness that additionally coincides with a ductile-to-brittle transition (DBT) in the polymer.
        To understand this DBT, we develop an elastic-plastic-damage finite element (FE) model with properties fit to the experimental strength and toughness data.
        We then explore how changes in both sample size and yielding zone size induce size-affected changes in fracture behavior.
        
        \begin{figure*}[ht!]
                \includegraphics[width=\textwidth]{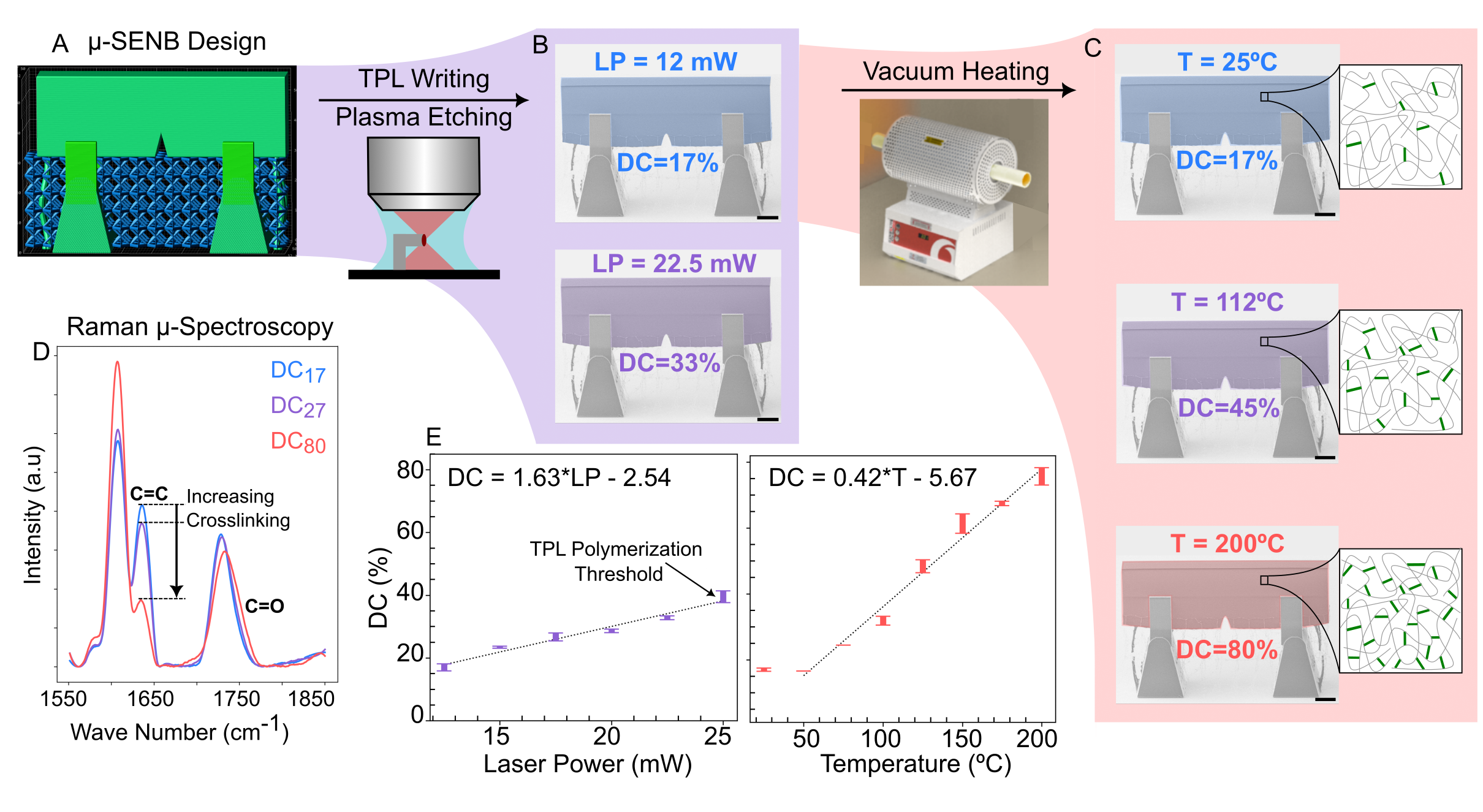} 
                \caption{\textbf{Polymer Degree of Conversion (\DC{}):}
                A) Graphic rendering of the \uSENB{} design.
                B) Specimens with \DC{} values between 17-39\% generated using different laser exposures in TPL. 
                C) Specimens with starting \DC{} value = 17\% are heated under vacuum to generate specimens with \DC{} values between 17-80\%. Insets illustrate increasing cross-linking density.
                D) Raman spectra for specimens with three different \DC{} values showing the decreasing intensity of the C=C peak, which correlates to increased cross-linking between the polymeric chains.
                E) Raman data showing the effect of laser power and temperature on \DC{}.
                All scale bars = 10\um{}.}
                \label{fig:Figure1} 
        \end{figure*}

    \section{Sample Preparation and Characterization}
        Microscale beams with varying degrees of conversion were designed in a microscale single-edge notch bend (\uSENB) configuration and fabricated out of a commercial acrylate-based photopolymer IP-Dip using TPL-DLW.
        Two separate variables were used to change the polymer degree of conversion: i) laser power was varied between 11 mW to 25 mW in $\sim$2.5 mW increments, and ii) annealing temperature was varied between 25°C to 200°C in $\sim$25°C intervals (Figure \ref{fig:Figure1}A,B,C and Figure S1).
        The maximum temperature of 200°C was chosen because it is well below the degradation onset temperature of the photopolymer of $\sim$250°C \cite{bauer2020thermal}.
        Using temperature as a control variable not only allows samples to have a higher \DC{} than what can be achieved with TPL alone, but it creates an overall more homogeneous material with a uniform degree of conversion through the cross-section.
        This minimizes any additional toughening due to material heterogeneity effects \cite{patel2023toughness}.

        Beam dimensions were determined using ASTM E-1820b \cite{ASTM-E1820} with a standard beam thickness of W = 26 \um{} and span-to-thickness ratio S/W=2.
        Notches were directly written into the beams during the printing process to avoid introducing any focused ion beam (FIB) milling damage \cite{ast2019review}; these had an initial length $a_{o}$ = 4.5 \um{} and a/W = 0.17.
        Additional geometrically scaled specimens with thicknesses equal to W, 2W/3, and W/3 were created to characterize the size of the fracture process zone and the material fracture energy.
        Displacement-controlled in-situ \uSENB{} fracture tests with continuous stiffness measurement (CSM) were conducted using a piezo-driven nanoindentation system (ASA, Alemnis AG).
        Specimens were tested either to complete fracture or to a displacement of 18 \um{}, and video data were captured for each test.

        Raman micro-spectroscopy performed on printed beam samples revealed that \DC{} increases linearly with both laser power (LP) and temperature (T) as $DC = 1.629*LP-2.54$ and $DC = 0.418*T-5.67$, respectively (Figure \ref{fig:Figure1}D,E and Figure S2).
        Varying the laser power produced \DC{} values between 17\% - 40\%, with 40-45\% being the upper limit that can be achieved without overexposing the resin.
        Heating of samples printed with \DC{}=25\% had no effect up to 50°C, after which the \DC{} linearly increased to a maximum of 80\% at 200°C.
        The mechanical properties of this resin system can be estimated based on literature data for both as-written samples \cite{bauer2019programmable} and for fully cross-linked samples \cite{bauer2020thermal}, which show a Young's modulus and yield strength variation between 1.5-4.3 GPa and 30-80 MPa, respectively, for the range of \DC{} studied here.
        Further details on the fabrication and testing procedures in this work can be found in \textit{Materials and Methods}.

        \begin{figure*}[ht!]
                \includegraphics[width=\textwidth]{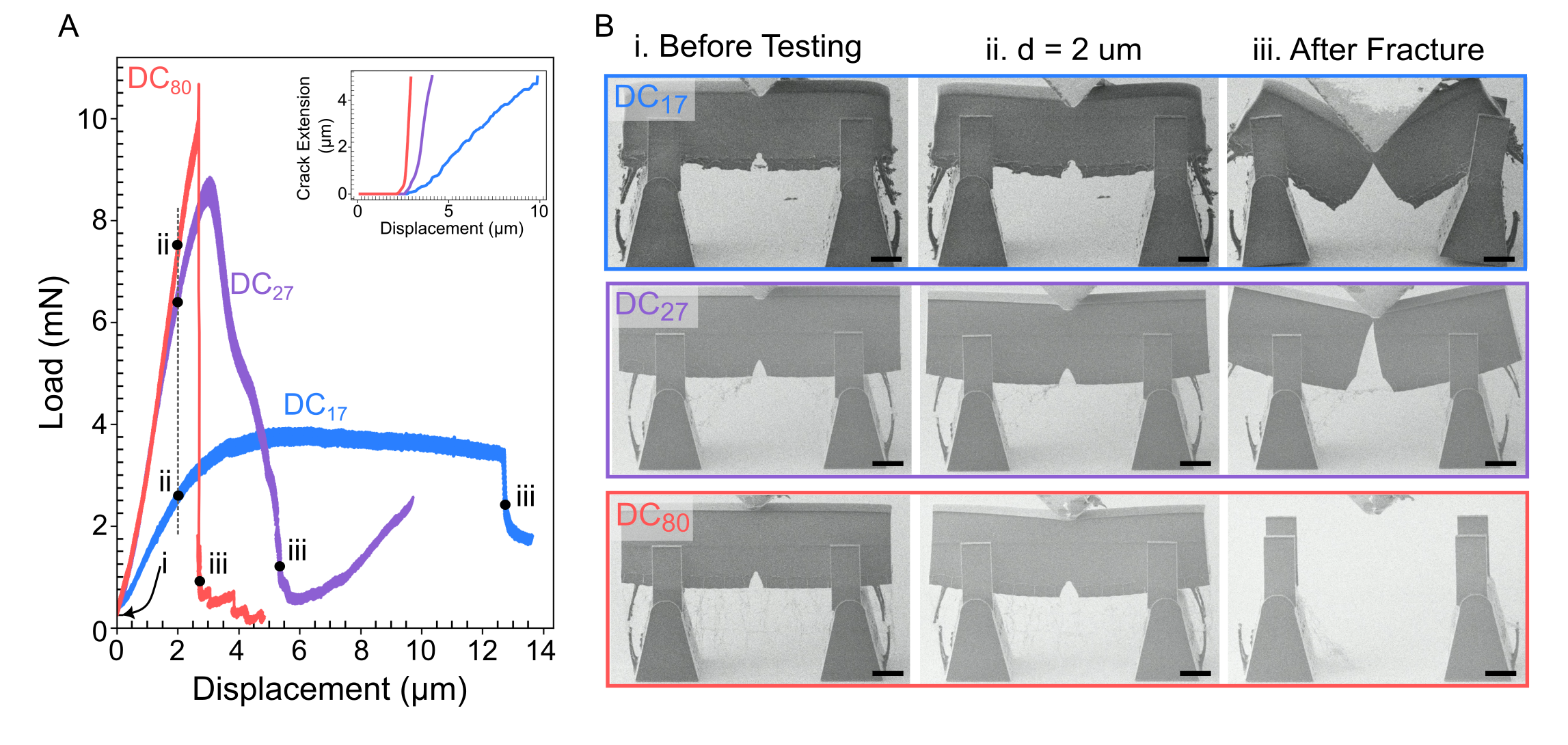} 
                \caption{\textbf{Mechanical Testing Results:}  
                A) Representative load-displacement curves for samples with \DClow{}, \DCmed{}, and \DChi{} with inset showing corresponding crack extension vs. displacement.
                Hollow circles represent the point of crack initiation. 
                B) SEM stills from in-situ videos showing the different failure behaviors observed. Scale bar = 10\um{}.}
                \label{fig:Figure2} 
        \end{figure*}

    \section{Stiffness, Strength and Fracture Energy}
        The mechanical response of the different test specimens was found to correlate strongly with the \DC{} regardless of whether samples were tested in an as-written state or were thermally annealed. 
        The peak load and bending stiffness both increased with \DC{}, ranging from 4 mN to 11 mN and 1.2 kN/m to 4.3 kN/m, respectively.
        There was little variation in strength or stiffness after $DC\approx$ 40\%, but specimens with a higher \DC{} showed significantly lower strains to failure.
        Representative load-displacement data are shown in Figure \ref{fig:Figure2}, and the complete data set for all samples tested is provided in Figure S3.

        We observed three characteristic deformation regimes depending on the \DC{} value, which we will distinguish here as \DClow{} for values between 17-25\%, \DCmed{} for values between 25-40\%, and \DChi{} for values greater than 40\%.
        Beams with \DClow{} showed elastic-plastic behavior with long plastic plateaus and slow, stable crack propagation.
        In \DCmed{} specimens, the mechanical behavior was linear up to peak load, followed by gradual softening and stable crack propagation. 
        In contrast, all \DChi{} specimens showed a linear-elastic behavior up to the peak load, followed by unstable crack propagation and catastrophic failure. 
        Stills from in-situ videos are shown in Figure \ref{fig:Figure2}, demonstrating the significant change in failure behavior with increasing \DC{} (see Supplementary Movies S1-S3).

\begin{figure*}[ht!]
                \includegraphics[width=\textwidth]{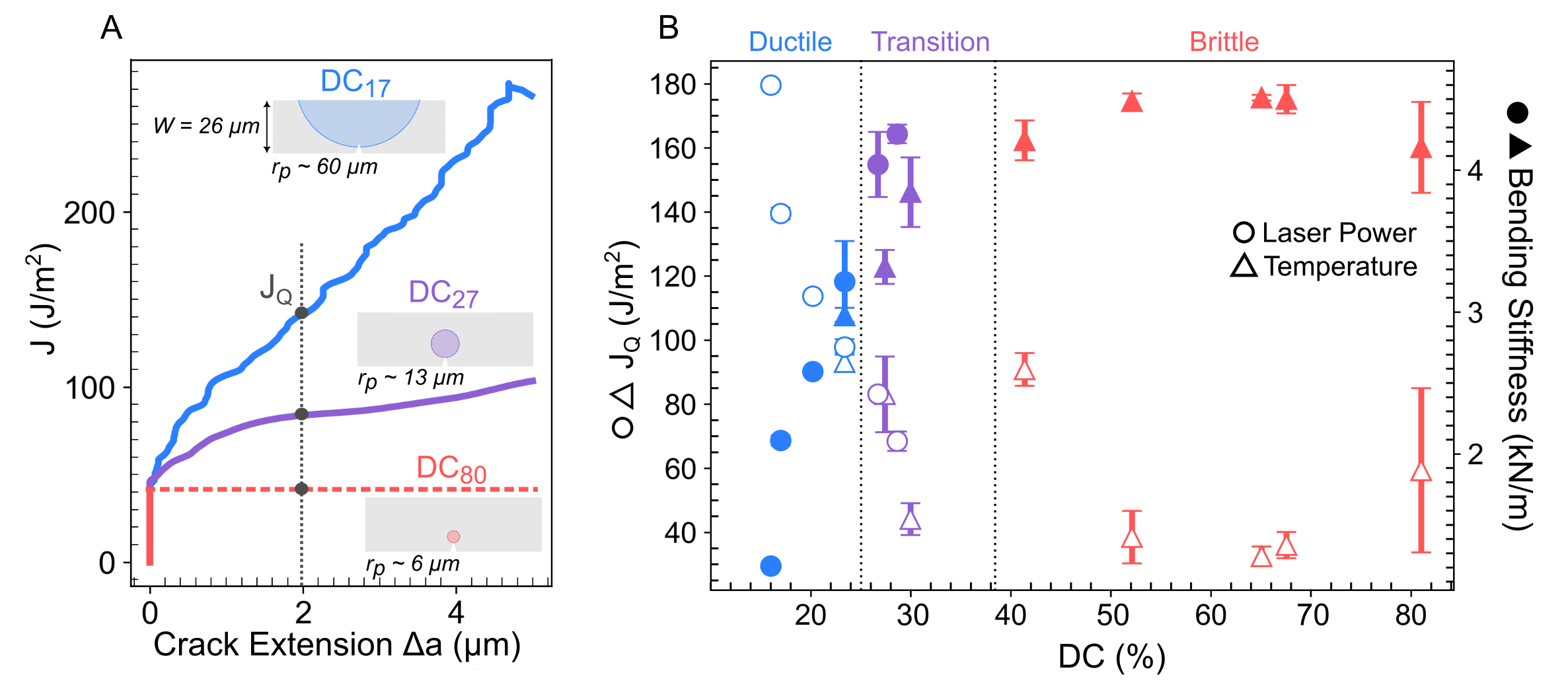} 
                \caption{\textbf{J-R curves, Fracture Energy and Bending Stiffness:} 
                A) Representative J-R curves showing decreasing crack growth resistance with increasing \DC{}. 
                Insets show plastic zone sizes of respective materials. 
                B) Fracture Energy (hollow markers) and bending stiffness (solid markers) vs \DC{} for all samples tested in this study, showing decreasing fracture energy and changing fracture behavior with increasing sample stiffness. 
                Error bars represent standard deviation values from at least three runs on each specimen type.}
                \label{fig:Figure3} 
        \end{figure*}

        The load-displacement and CSM data were further analyzed to quantify the fracture properties, details of which can be found in \textit{Materials and Methods}.
        Three distinct J-R crack resistance behaviors were observed across the range of \DC{} values studied, as shown in Figure \ref{fig:Figure3}. 
        In the \DClow{} samples, the moderate softening after the peak load resulted in stable crack propagation and rising J-R curves, with fracture energies as high as \Jq{} = 180\Jms{}, where \Jq{} is the J value at 2\um{} of crack extension.
        In samples with \DCmed{}, the greater post-yield softening caused a faster crack propagation, resulting in lower fracture energies of \Jq{} = 40-60\Jms{}.
        The lower J-R curve slope and near plateau in the J value of the \DCmed{} samples indicate that they have a nearly fully developed yielding zone.
        For samples with \DChi{}, crack growth was unstable, and they had correspondingly lower fracture energies of \Jq{} = 30\Jms{}.
        This pronounced $\sim$6x reduction in \Jq{} between \DC{} = 17\% and \DC{} = 80\% specimens is shown in Figure \ref{fig:Figure3}, illustrating that the enhanced strength and stiffness from thermal treatment comes at a significant cost of toughness.
        These results demonstrate that an optimal balance between strength and toughness can be obtained for samples with a \DC{} in the range of 20-25\%.
        Note that the \Jq{} values here should not be considered true material properties for any test where the J-R curve did not plateau, but they do represent meaningful trends in material properties.
        
        The large drop in fracture energy with increasing \DC{} in this work matches well with existing models for polymer fracture energy, which is known to depend strongly on the proportions of cross-linked and entangled chains. 
        Materials with high chain cross-linking density tend to experience brittle failure via chain scission, while those with high chain entanglement density undergo ductile failure via chain sliding \cite{zhurkov1974atomic, wang2014phenomenological, dookhith2022tailoring}.
        Various models have been developed to assess the impact of network architecture and defects on fracture energy (\Gt{}) \cite{lake1967strength, arora2020fracture, lin2021fracture}, which find that fracture energy scales inversely with chain cross-linking density ($v_x$) as \Gt{} $\propto v_{x}^{-1/2}$ in highly elastic materials and \Gt{} $\propto v_{x}^{-4}$ in elastic-plastic materials \cite{dookhith2022tailoring}.
        In both cases, lower cross-linking density leads to greater energy dissipation and a corresponding enhanced toughness.
        However, despite attempts to connect this fracture energy to characteristic length scales \cite{dookhith2022tailoring, brown1982model}, there is no mechanistic explanation for the observed ductile to brittle transitions in this or other work.
        
    \section{Ductile-to-Brittle Fracture Transition}
        Ductile-to-brittle transitions are well known to occur in various crystalline and amorphous polymers \cite{brown1982model, quagliato2022quasi, argon2003toughenability}. 
        These transitions have been connected to external factors such as temperature and loading rate and internal factors like molecular weight, density, and microstructure \cite{brown1982model}. 
        One essential aspect of these transitions is the ratio of sample size ($D$) to characteristic fracture length scales.
        Prior to fracture, a yielding zone of size \rp{} will develop in front of a flaw. 
        This encompasses a fracture process zone (FPZ) of size \lch{}, i.e., a damage-driven softening zone at the crack tip, and a plastic zone (PZ) of size (\rp{}--\lch{}), i.e., a plastic hardening zone around the FPZ.
        Standards for measuring fracture toughness generally require specimens to be significantly larger than these fracture length scales (i.e., $D \gg$ \rp{}) to ensure a small-scale yielding condition \cite{ASTM-E1820}.
        In these scenarios, samples will form a small plastic region at the tip of a crack while the rest of the sample remains linear elastic, and brittle failure -- i.e., unstable crack growth -- will occur when a crack or flaw reaches a critical size.

        \begin{figure}[ht!]
            \centering
            \includegraphics[width=8.5cm]{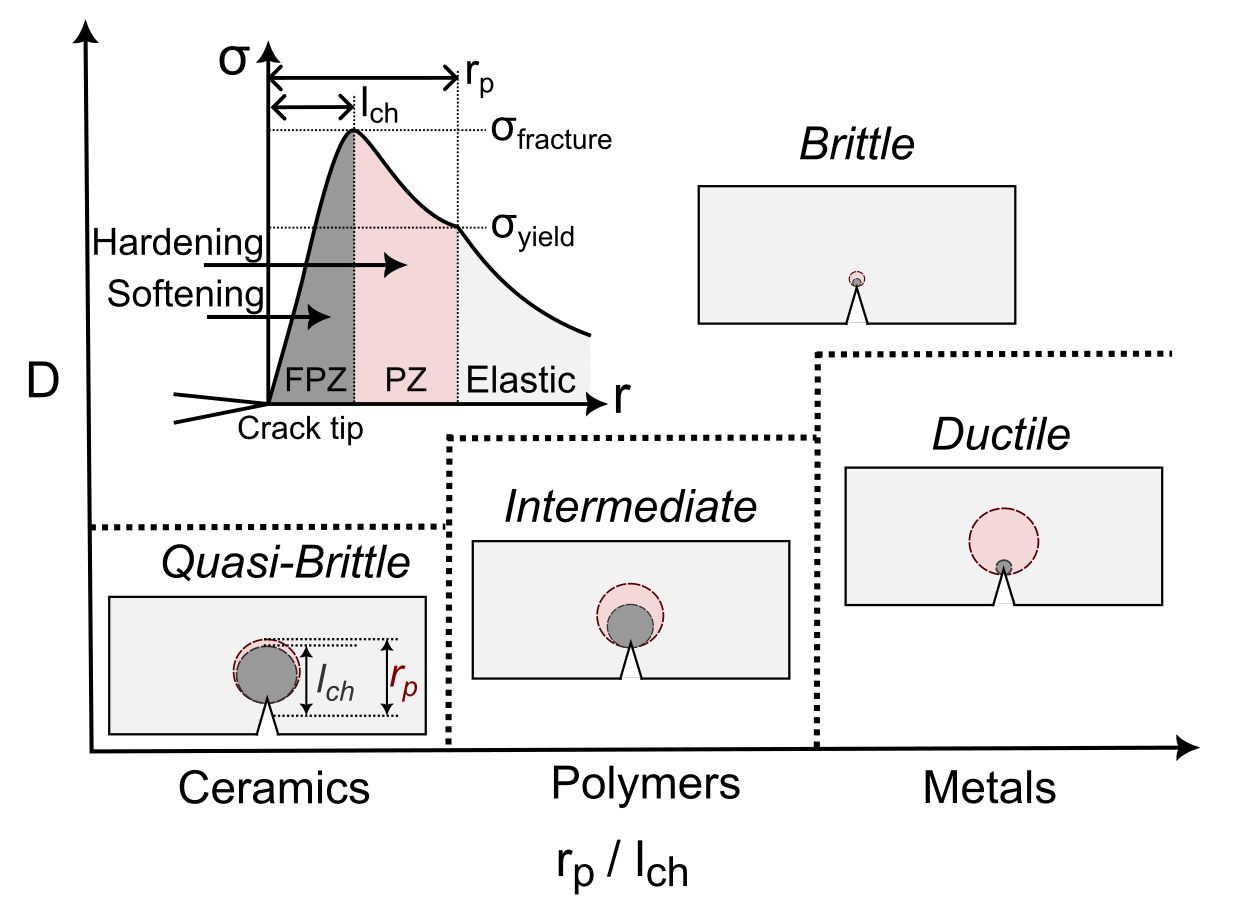}
            \caption{\textbf{Fracture Transitions} 
            Changing fracture behavior with changing sample size ($D$) w.r.t to the relative size of the sample yielding zone (\rp{}) and fracture process zone (\lch{}).}
            \label{fig:Figure4}
        \end{figure}
        
        Small samples with $D \leq$ \rp{} generally have an enhanced ductility due to energetic size-effects, namely when there is insufficient fracture energy to cause unstable crack growth according to linear elastic fracture mechanics (LEFM).
        When the PZ is much larger than the FPZ (i.e., \lch{} $<<$ \rp{}), as is the case for many metals, materials will exhibit an increasingly ductile response due to large scale yielding as $D$ is decreased \cite{nguyen2021structural}.
        When the FPZ is much larger than the PZ (i.e., \lch{} $\approx$ \rp{}), materials will show a quasi-brittle response wherein a large damage zone will develop that suppresses crack propagation, as is the case for most ceramics \cite{bavzant1990size}, concrete \cite{bavzant1984size} and polymer nanocomposites \cite{mefford2017failure}.
        An understanding of ductile-to-brittle transitions is currently lacking for polymeric material systems where both PZ and FPZ can be significant and, at times, comparable.
        Irrespective of the material and size-effect, samples that are sufficiently large ($D \gg$ \rp{}) will undergo fracture-driven failure, while samples with sizes below the characteristic fracture length ($D \leq$ \rp{}) will experience strength-driven failure \cite{bavzant1984size, bavzant1990size, bavzant2000size, ba1990determination, ast2019review, qiao2019strength}.
        This is illustrated in Figure \ref{fig:Figure4}.
        
        Energetic fracture size-effects can be analyzed using Ba\v{z}ant-type size effect laws, which quantify changes in nominal strength ($\sigma_{N}$) with sample size.
        The size-effect law (SEL) was initially developed for quasi-brittle materials where FPZ $\gg$ PZ \cite{bavzant1984size}.
        More recently, Nguyen et al. \cite{nguyen2021structural} have developed a model for materials whose PZ $>>$ FPZ, which as a first approximation, is more representative of the materials studied here.
        This SEL law is expressed as:
        
        \begin{align}
            \sigma_{N} = \frac{\sigma_{o}}{\sqrt{1+D/D_{o}}},  \label{Eq:SEL}
        \end{align}
        where
        \begin{align}
            \sigma_{o} = \sqrt{E^*G_{t}/2r_p}, \text{ and }  D_{0} = 2r_p/g(\alpha_{0}). \label{Eq:SEL2}
        \end{align}
        
        Here, $\sigma_{o}$ is the maximum material strength, $E^*$ is the effective Young's modulus, \Gt{} is the total fracture energy,  $g(\alpha_o)$ is the dimensionless energy release rate, \rp{} is the size of the yielding zone, and $D_{o}$ is the characteristic fracture length scale \cite{guinea1998stress, nguyen2021structural}.
        In the limits of this model, materials with small dimensions ($D << D_o$) will experience strength-governed failure as $\sigma_{N} \propto D$, while materials with large dimensions ($D >> D_o$) will experience fracture-driven failure as $\sigma_N \propto D^{-1/2}$.
        Despite the extensive work in this area, there is still a need for better theoretical frameworks when FPZ $\sim$ PZ.
        More importantly, it is crucial to characterize the relative process zone sizes to understand the underlying causes of ductile to brittle transitions.
        
        \subsection{Yielding zone estimation}
        We first assess the size of the yielding zone using a simple LEFM approximation of $r_p = \frac{1}{2\pi}\frac{G_{t}E^*}{\sigma_y^2}$, where $\sigma_y$ is the yield strength.
        It should be noted that this is a generalized expression that does not account for the plastic hardening or the geometry of the structure, both of which can significantly impact the plastic zone \cite{kudari2007effect}.
        Using the approximate fracture energy dissipation \Jq{} measured in this study and yield strength values from literature \cite{bauer2019programmable, bauer2020thermal}, we estimate \rp{} values of $\sim$60\um{}, 13\um{}, and 6\um{} for \DCi{}, \DCj{} and \DCk{}, respectively, where, e.g., \DCi{} is a DC of 17\%.
        It is apparent that changes in cross-linking cause a significant change in process zone size, with \rp{} decreasing by $\sim$10x as \DC{} changes from 20-80\%.
        The \rp{} values are illustrated with respect to the beam dimensions in Figure \ref{fig:Figure3}A.

        \subsection{SEL analysis} \label{SEL analysis}
        An SEL analysis was used to obtain a more precise experimental yielding zone size by taking specimens of varying sizes and analyzing changes in their nominal strength with sample size, as described by equations \ref{Eq:SEL} and \ref{Eq:SEL2}.
        Three geometrically scaled \DCj{} and \DCk{} samples with beam thicknesses of 8.5\um{}, 17\um{} and 26\um{} were tested to failure, and their load-displacement data was analyzed using the SEL-based linear regression method to determine \Gt{} and \rp{} \cite{nguyen2021structural}.
        Details of the calculations can be found in \SI{} (Figure S4).
        \DCi{} samples were excluded because the estimated \rp{} is larger than what could realistically be made using TPL-DLW, meaning the SEL analysis would be invalid.
        \Gt{} values of $\sim 68 J/m^2$ and $\sim 55 J/m^2$ and \rp{} values of 8\um{} and 2\um{} were obtained for \DCj{} and \DCk{}, respectively. 
        These results demonstrate that the \rp{} size is reduced by a factor of $\sim$4x as DC increases from 27\% to 80\%, and importantly that the \rp{} for the \DCj{} specimens begins to approach the 26\um{} sample thickness.

        \begin{figure*}[ht!]
                \includegraphics[width=\textwidth]{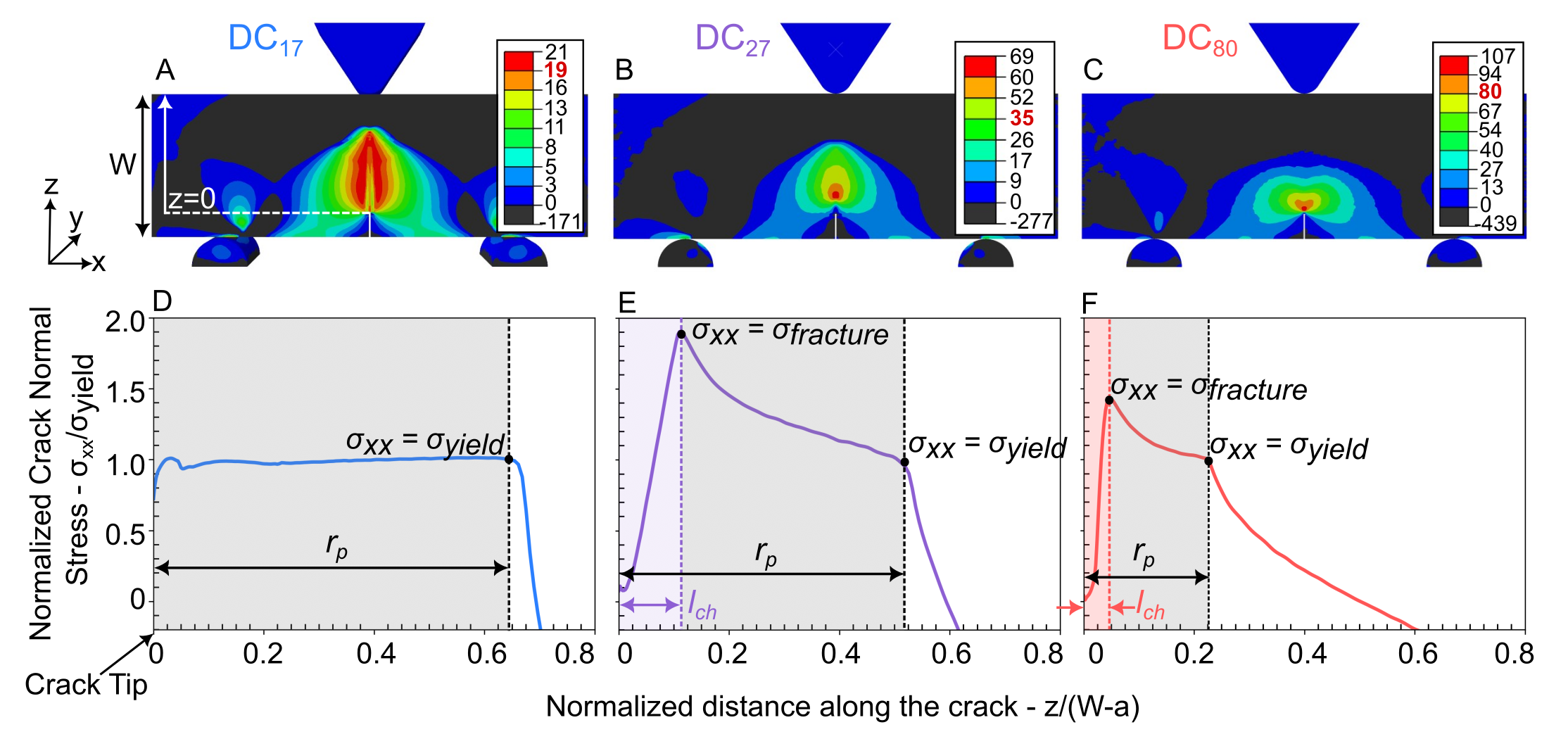}
                \caption{\textbf{Numerical modeling of FPZ and PZ} 
                A, B, C: Crack normal stress ($\sigma_{xx}$) distribution in the mid-plane of \DCi{}, \DCj{} and \DCk{} W=26\um{} beams, after the yielding zone is fully developed.
                The bold-red value in the scale bar indicates the yield strength. 
                D, E, F: Crack normal stress normalized by yield strength as a function of distance from the crack tip normalized by length (W-a) for \DCi{}, \DCj{} and \DCk{} beams, respectively.  
                The plots highlight the yielding zone ($r_p$) and the relative size of the FPZ (\lch{}).}
                \label{fig:Figure5}
        \end{figure*}
        
        \subsection{Numerical quantification of $r_p$ and $l_{ch}$}
        An elastic-plastic-damage FE model \cite{salviato2022adding} was implemented to more thoroughly investigate the yielding zone size and shape, along with the relative sizes of the constituent FPZ and PZ with changing \DC{}.
        Model properties were fit directly from experimental data, and full details of the model setup are provided in \textit{Materials and Methods} and in the \SI{}.
        This model allows an estimate of the maximum possible FPZ and PZ size, as well as the fracture energy due to damage (\Gf{}). 
        
        In this model, the process zones are considered to be fully developed when i) the first element at the crack tip reaches a stress-free state and ii) the stress gradient does not change as the crack propagates.
        The crack normal stress ($\sigma_{xx}$) at this instant is plotted against the distance from the crack tip in Figure \ref{fig:Figure5}.
        The value of \lch{} is taken to be the distance from the crack tip to the point of maximum stress or fracture stress ($\sigma_{fracture}$).
        In \DCj{} and \DCk{} samples, the \lch{} values are found to be 2.4\um{} and 1.0\um{} respectively.
        In the \DCi{} samples, the small sample size did not allow the FPZ to fully develop, as indicated by the non-zero stress at the tip of the crack (Figure \ref{fig:Figure5}D).
        To determine \lch{}, a 120x scaled-up version of the beam was modeled, revealing an FPZ size of 115\um{} (Figure S6).

        The PZ size is taken to be the distance from the peak stress in front of the crack tip (i.e., the end of the FPZ) to the point where $\sigma_{xx}=\sigma_y$ (Figure \ref{fig:Figure5}).
        In the 26\um{} thick beams, plastic zones were found to have a considerable size of 14\um{}, 11\um{}, and 5\um{} for the \DCi{}, \DCj{}, and \DCk{} samples, respectively.
        It should be noted that the PZ was not fully developed for the \DCi{} sample here, and in 120x scaled simulations for the \DCi{} beam, the fully developed PZ size is found to be 165\um{}. 
        
        These results provide critical insight into how fracture behaviors change according to FPZ and PZ size.
        As the \DC{} decreases from 80\% to 17\%, the FPZ and PZ size increase by $\sim$115x and $\sim$33x, respectively.
        This substantial size increase is most prominent in the \DClow{} specimens, and the change is more modest between the \DCmed{} and \DChi{} specimens.
        These trends align with observed variations in strength and bending stiffness, which show a similarly steep increase in the \DClow{} regime (Figure \ref{fig:Figure3}).
        Importantly, these results demonstrate that for the \DCi{} beams, both the FPZ and PZ are significantly larger than the 26\um{} experimental sample thickness.
        This indicates that the observed ductility in the \DClow{} specimens is the result of an underdeveloped yielding zone that does not meet the \Gt{} required to cause fracture.

        \begin{table} [h!]
            \centering
            \caption{Damage fracture energy (\Gf{}), plastic fracture energy (\Gp{}), FPZ (\lch{}) and yielding zone (\rp{}) for different degrees of conversion (DC)}
            \begin{tabular}{ccccc}
                DC  &   \Gf{} (\Jms{})  &   \Gp{} (\Jms{})  &   \lch{} (\um{})  &   \rp{} (\um{})\\
                \midrule
                17  &       90          &       -           &       115         &       280 \\
                27  &       28          &       40          &       2.4         &       11 \\
                80  &       18          &       37          &       1.0         &       5 \\
                \label{Table1}
            \end{tabular}
        \end{table}
        
        \subsection{Damage and plasticity governed fracture}
        The plastic energy dissipation rate \Gp{} was determined by subtracting the damage fracture energy \Gf{} input for the FE model from the total fracture energy \Gt{} determined from the SEL experiments (i.e., \Gp{} = \Gt{} - \Gf{}), the results from which are shown in Table \ref{Table1}.
        We find that in the \DCk{} specimens, the plastic energy dissipation rate (\Gp{}=37\Jms{}) contributes $\sim$2x that of the energy dissipated via damage (\Gf{}=18\Jms{}), despite exhibiting very brittle failure.
        In comparison, despite an almost comparable plastic energy dissipation rate (\Gp{} = 40\Jms{}), the \DCj{} specimens have a slightly higher total fracture energy, mainly due to an increase in the damage energy dissipation (\Gf{} = 28\Jms{}).
        The \DCi{} specimens have a very high damage energy dissipation of \Gf{}=90\Jms{}, and likely have an even higher \Gp{}, but it was not possible to estimate this value because SEL experiments were not conducted on these specimens.
        These relative changes in plastic and damage energy dissipation correlate well with the increased \rp{} and \lch{} for lower \DC{} samples.
        
        It is interesting to note that the PZ size is larger than the FPZ size across all the \DC{} values examined in this work.
        This, along with the consistent trend of \Gp{}$>$\Gf{} for all the samples, indicates that the fracture behavior is more affected by plasticity than damage.
        However, this difference is not substantial enough to cause either ductile or quasi-brittle behavior, as illustrated in Figure \ref{fig:Figure5}.
        The fracture behavior of these photopolymers thus falls in an intermediate regime, where both softening and hardening ahead of the crack tip impact crack growth, warranting further analysis to understand the emergent size-affected behavior.
        This also suggests that the \rp{} and \Gt{} values calculated using the SEL law in equation \ref{Eq:SEL} and \ref{Eq:SEL2} in section \ref{SEL analysis} might not be fully accurate and there is a need for improved SEL models that can capture this behavior more comprehensively. 
        
        The yielding zone sizes determined from the numerical model are comparable to the estimates obtained from the LEFM model for the \DCmed{} and \DChi{} specimens, but the \DClow{} LEFM estimate significantly underestimated the actual \rp{} size.
        This result is unsurprising given that the \Jq{} value obtained from the experiments is an underestimate of the actual \Gt{}, but that the \DCmed{} and \DChi{} specimens had a nearly fully developed FPZ.
        The LEFM estimate nevertheless provided a quick and useful estimation of the characteristic material length scale that would be useful for predicting size-affected changes in properties.

        \subsection{Size-affected ductility}
\begin{figure}[h!]
            \centering
                \includegraphics[width=8.5 cm]{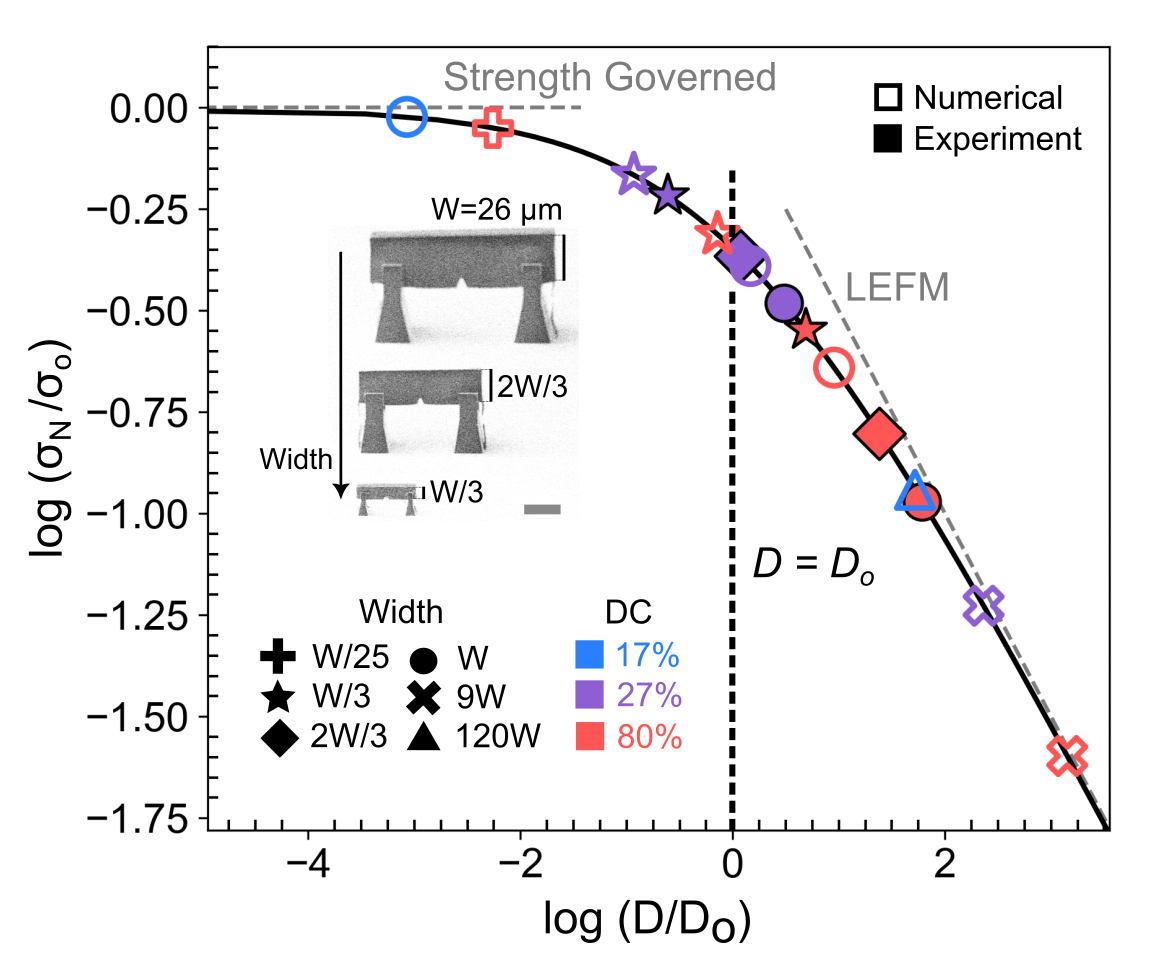}
                \caption{\textbf{LEFM vs. strength governed failure.} 
                A SEL plot of nominal strength ($\sigma_N$) normalized by material strength ($\sigma_o$) vs. sample dimension ($D$) normalized by the transition fracture length ($D_o$). Experimental and numerical results are shown for different sample sizes and degrees of conversion. Scale bar: 10 $\mu$m.}
                \label{fig:Figure6}
        \end{figure}

        We now evaluate the change in sample strength ($\sigma_N$) with sample size ($D$) in the context of an SEL analysis.
        The parameters to calculate the constituent material strength ($\sigma_o$) and transition fracture length ($D_o$) (equation \ref{Eq:SEL2}), are determined from SEL experiments in combination with numerical models, and normalized SEL data is shown in Figure \ref{fig:Figure6}.
        Comparing the sample strength w.r.t to the material strength ($\sigma_N/\sigma_o$) for structures of the same size ($D=W=$26\um{}) revealed that decreasing the \DC{} causes a transition from flaw-based to strength-based failure.
        
        For the \DCk{} specimens, the yielding zone is significantly smaller than the sample size ($D\gg$ \rp{}), causing it to fall close to the LEFM-dominated regime of the SEL curve.
        In the \DCj{} specimens, the yielding zone size approaches the sample size but is still smaller ($D<$\rp{}), causing an observed transition from LEFM to strength-based fracture.
        The \DCi{} specimens are significantly smaller than the yielding zone size ($D<<$ \rp{}), causing it to fall well into the strength-governed fracture regime.
        It is clear from these trends that \textit{ductility emerges when $D \sim$ \rp{}}.

        These results and additional results from specimens of sizes of 2$W$/3 and $W$/3 are shown to fit well onto the SEL curve.
        Experimentally scaled down \DCj{} and \DCk{} specimens do not show the same pronounced ductility as the \DCi{} specimens, but they do show a slower crack propagation velocity (Figure S7).
        Numerical models on a wider range of specimens reveal changes in failure behavior for samples with the same \DC{}.
        The 120$W$ \DCi{} specimens show a characteristic brittle failure, while the $W$/25 \DCk{} specimens undergo a similar ductile failure to the \DCi{} specimens (Figure S6).
        It should be noted that the slight discrepancy between the experimental and numerical results on the SEL curve is the result of different methods used to calculate \rp{}, namely the SEL method for experiments and a direct measurement for the FE models, again reinforcing the need for better SEL models for this intermediate fracture regime.
        These results illustrate the utility of simple size-effect laws in understanding the changes in fracture behavior.
        Despite this, the need for a full numerical model to quantify fracture energy and FPZ/PZ size highlight the complex interplay between structural dimensions and material properties and reestablish the need for improved size-effect models for when FPZ$\sim$PZ.

    \section{Summary and Perspectives}
        In this work, we use \uSENB{} experiments and corresponding FE modeling to study the fracture behavior and process zone sizes of a TPL-DLW polymer as a function of \DC{} and part size. 
        While increasing the \DC{} from 17\% to 80\% increases both strength and stiffness by $\sim$3x and $\sim$4x, respectively, there is a corresponding $\sim$6x reduction in toughness and a corresponding DBT in the failure behavior.
        Experiments on differently sized beams in combination with an elastic-plastic-damage FE model reveal significant changes in fracture energy and yield zone size \rp{}, which change by nearly two orders of magnitude across the \DC{} range studied here.
        Despite this variation, the relative sizes of the FPZ and PZ were comparable, indicating that the observed fracture behavior falls in an intermediate range between ductile and quasi-brittle fracture.
        By analyzing the specimens in the context of an SEL framework, it becomes apparent that ductility emerges when the sample size approaches the yielding zone size ($D\approx$\rp{}) and additionally corresponds with a transition from flaw-governed to strength-governed failure.
        These results demonstrate that ductility is an emergent property even in traditionally brittle materials and that process parameters and feature size are both crucial factors in creating materials with enhanced mechanical properties.

        These results clearly demonstrate that understanding characteristic fracture lengths is crucial for designing any component where mechanical performance is a priority. 
        Processing or compositional changes such as heating, filler addition, or grain size modification inevitably alter process zone sizes and can drastically alter fracture behaviors.
        In the context of architected materials and composites, there is a new potential to design feature sizes to match constituent characteristic length scales and maximize toughness while maintaining strength and stiffness.
        The strength and stiffness of a structure can be controlled by altering the material composition, while the fracture behavior can be modified by adjusting the feature sizes within the architecture. 
        
        In the case of parts produced using TPL, the typical minimum feature sizes of 200nm - 5\um{} are in the same range, and the process zone sizes of many of the constituents, making it a perfect vehicle for studying size-affected changes in architected material toughness.
        Importantly, greater attention needs to be paid to materials with a high \DC{}.
        While thermal treatment has been suggested as a solution to property variation between prints \cite{bauer2020thermal}, the resulting small process zone leads to more brittle behaviors, as is likely the case in materials that have inadvertently been thermally annealed during atomic layer deposition or plasma etching \cite{meza2015resilient, moestopo2023knots}.
        Considering the role of feature size on ductility in any additive manufacturing material can offer enhanced tunability in structural design.
        It additionally becomes possible to cascade multiple damage mechanisms in hierarchical architectures to develop highly tough and damage-tolerant materials, as is commonly seen in many natural structural materials \cite{huang2019multiscale}.

    \section*{Experimental and Numerical Methods}
    \subsection*{Fabrication} \label{Materials and Methods}
        The entire three-point bend assembly was designed in Python and imported into the Photonic Professional DLW system (Nanoscribe GmbH).
        Specimens were fabricated on silicon substrates, which were first etched for 5 minutes in oxygen plasma (Plasma Etch PE25) and then functionalized using 3-(Trimethoxysilyl) propyl methacrylate to improve adhesion and prevent peeling off of supports during fracture tests. 
        Printing was done using two-photon lithography (TPL) direct laser writing (DLW) system (Nanoscribe, GmbH).
        A proprietary acrylate-based resist, IP-Dip (Nanoscribe, GmbH), was used with a 63x objective to achieve high-precision, sub-micron resolution writes. 
        Writing speeds, specimen hatching, and layering were all kept constant at 10 mm/sec, 100 nm, and 300 nm, respectively, to minimize any heterogeneity from the TPL-DLW process
        Support lattices were written in piezo mode with the lowest laser power (6 mW) to ensure faster etching. 
        After printing, samples were immersed in a propylene glycol monomethyl ether acetate (PGMEA) solution for 20 mins, then in ultrapure IPA for 30 mins, followed by critical point drying (Tousimis Autosamdri-931). 
        These were subsequently etched in an oxygen plasma etcher (YES CV200 RFS) at 65 W power for 25-35 mins until the support lattices were completely removed, producing free-standing fracture specimens.
        This support lattice with plasma etching method is similar to one developed by Gross et al. \cite{gross2019additive}. 
        Thermal treatment samples were written with a 15mW laser power, then heated in a high vacuum tube furnace (Carbolite Gero), and maintained at their peak temperature for 1 hour.
        Large sweeps were written to quantify the polymer degree of conversion as a function of laser power and temperature (Figure S1).

    \subsection*{Raman Micro-spectroscopy}
        Raman Spectra were acquired using an inVia (Renishaw plc) confocal Raman microscope with a 50× objective, operated at an excitation wavelength of 785 nm, with a laser intensity of 50\% and an exposure time of 10s averaged over 3 acquisitions.
        DC values were extrapolated from Raman spectra using the relationship 
        $DC = 1 - (\frac{A_{C=C}/A_{C=O}}{A^{'}_{C=C}/A^{'}_{C=O}})$ \cite{baldacchini2009characterization}.
        Here, $A_{C=C}$ and $A_{C=O}$ are the integrated intensities of the carbon-carbon and carbon-oxygen double bond peaks in the polymerized resin, respectively and $A^{'}_{C=C}$ and $A^{'}_{C=O}$ are the integrated intensities of the same peaks in the unpolymerized resin.

    \subsection*{Nanomechanical Testing}
        Fracture tests were performed using the displacement-controlled mode in an in-situ nanoindentation system (Alemnis ASA).
        Testing was done in a scanning electron microscope (SEM) (Thermo-Fisher Scientific Apreo) with a 2\um{} radius conductive diamond wedge tip. 
        A loading rate of 20 nm/sec was used, and beams were taken up to failure or 18\um{} piezo displacement. 
        A sinusoidal signal of amplitude 40 nm and frequency 4 Hz was superimposed to perform continuous stiffness measurement (CSM) and thereafter compute instantaneous crack lengths. 
        To account for thermal drift, the nanoindenter assembly was installed in the SEM and allowed to stabilize before testing.
        We additionally added `out-of-contact' segments before and after the compression step where the tip is not in contact with the sample to correct for any drift in the system. 
        Thermal drift showed a linear correlation with time for short experiment times (less than 30 minutes) and was subsequently subtracted from the load data during data processing.
        An amplitude-based Fast Fourier Transform (FFT) noise filtering algorithm was used to remove noise from the load-displacement data. 
        Subsequent CSM Data was smoothed using the Savitzky-Golay filter by fitting a third-order polynomial for every 300-400 data points.

    \subsection*{J-R Curve calculation}
        Instantaneous load line stiffness was calculated using the unloading slope of the CSM data, and crack initiation was determined as the point where the unloading stiffness began to decrease.
        Crack lengths were obtained using a compliance calibration procedure \cite{haggag1984compliance} by correlating the crack initiation point with the instantaneous stiffness thereafter.
        An elastic-plastic Mode-I J-integral was used to determine the samples' fracture behavior and crack growth resistance as defined in ASTM E1820-20b \cite{ASTM-E1820}.
        In softer samples, a critical J-integral value could not be computed using traditional EPFM methods since the J-R curves did not plateau. 
        We, therefore, define a conditional $J_Q$ value to be the J-integral value at a crack extension of $\Delta a=2$\um{}. 
        It should be noted that this $J_Q$ value is a severe underestimation of the true fracture energy for these materials.
        In samples that show brittle behavior and fail catastrophically, $J_Q$ was calculated at the point of maximum load.

    \subsection*{Computational Framework}
        To account for the significant plasticity and post-failure strain softening exhibited by the TPL polymers, an elastic-plastic-damage model was adopted from Salviato \cite{salviato2022adding} and implemented as a VUMAT subroutine in ABAQUS/Explicit with C3D8R mesh elements.  
        Beams were modeled to be homogeneous, i.e., without any cross-linking gradients, because the in-plane hatching and out-of-plane layering distances between the voxels were sufficiently smaller than the size of the voxel.
        Sharp cracks, with a crack tip radius comparable to the size in the fabricated beams ($\sim$ 250 nm), were used, and it was assumed that the cracks would propagate in a self-similar manner.
        The support structures and the indenter tip were modeled as elastic materials with an elastic modulus and Poisson's ratio of 3.5 GPa, 0.35, and 200 GPa, 0.33, respectively.
        A frictional penalty contact was defined between the beam and supports, as well as between the beam and the indenter tip.
        Materials were taken to be elastic-linear hardening with a tension/compression strength asymmetry and damage evolution as a function of equivalent plastic strain using a crack band model \cite{qiao2019strength, qiao2020micro}.
        Material properties were first estimated using uniaxial tension and compression data from Bauer et al. \cite{bauer2019programmable, bauer2020thermal} and then refined to fit experimental data in this work.
        The elastic modulus, yield strength, and plastic stress-strain properties were obtained from bend tests on unnotched beams, while the fracture energy was determined from experiments on notched beams (Figure S5).

    \section*{Acknowledgements}
    The authors gratefully acknowledge the financial support from the National Science Foundation under the Mechanics of Materials and Structures program managed by S. Qidwai (award no. 2032539).
    Part of this work was conducted at the Washington Nanofabrication Facility and Molecular Analysis Facility, a National Nanotechnology Coordinated Infrastructure (NNCI) site at the University of Washington, with partial support from the National Science Foundation via awards NNCI-1542101 and NNCI-2025489.

    \section*{Author Contributions}
    Ideation, Z.S.P., L.R.M.; experimental investigation, Z.S.P. and A.O.A.; numerical investigation, K.D, and M.S.; Writing – Original Draft, Z.S.P., L.R.M.; Writing – Review and Editing, all authors; Funding Acquisition, L.R.M.; Supervision,  L.R.M.

    \section*{Declaration of Interests}
    The authors declare no competing interests.
    
    \bibliographystyle{unsrt}
    \bibliography{References}

\end{document}